# Scattering properties of PT-symmetric chiral metamaterials

**Ioannis Katsantonis** [1,2,*], **Sotiris Droulias** [1,2], **Costas M. Soukoulis** [1,3], **Eleftherios N. Economou** [1,4] **and Maria Kafesaki** [1,2,*]

[1] Institute of Electronic Structure and Laser, Foundation for Research and Technology Hellas, 70013, Heraklion, Crete, Greece; sdroulias@iesl.forth.gr (S.D.)
[2] University of Crete, Department of Material Science and Technology, 70013, Heraklion, Greece
[3] Ames Laboratory and Department of Physics and Astronomy, 50010, Iowa States, Ames, USA; soukoulis@ameslab.gov (C.M.S)
[4] University of Crete, Department of Physics, 70013, Heraklion, Greece; economou@iesl.forth.gr (E.N.E)
[*] Correspondence: katsantonis@iesl.forth.gr (I.K.); kafesaki@iesl.forth.gr (M.K.)

**Abstract:** The combination of gain and loss in optical systems that respect parity-time (PT)-symmetry has pointed recently to a variety of novel optical phenomena and possibilities. Many of them can be realized by combining the PT-symmetry concepts with metamaterials. Here we investigate the case of chiral metamaterials, showing that combination of chiral metamaterials with PT-symmetric gain-loss enables a very rich variety of phenomena and functionalities. Examining a simple one-dimensional chiral PT-symmetric system, we show that with normally incident waves the PT-symmetric and the chirality-related characteristics can be tuned independently and superimposed almost at will. On the other hand, under oblique incidence, chirality affects all the PT-related characteristics, leading also to novel and uncommon wave propagation features, such as asymmetric transmission and asymmetric optical activity and ellipticity. All these features are highly controllable both by chirality and by the angle of incidence, making PT-symmetric chiral metamaterials valuable in a large range of polarization-control-targeting applications.

## 1. Introduction

A quantum mechanical Hamiltonian is called PT-symmetric, if and only if, it commutes with the combined action of parity, $\hat{P}$, and time-reversal, $\hat{T}$, operators, i.e. $[\hat{P}\hat{T}, \hat{H}] = 0$, with the brackets denoting the commutator. Since the action of parity on the momentum and position operators, $\hat{p}$ and $\hat{r}$ respectively, results to $\hat{p} \rightarrow -\hat{p}$, $\hat{r} \rightarrow -\hat{r}$, and the time-reversal to $\hat{p} \rightarrow -\hat{p}$, $\hat{r} \rightarrow \hat{r}$ and $i \rightarrow -i$ [1-4], the requirement for PT-symmetry in a one dimensional quantum Hamiltonian results to the symmetry condition for the potential $V^*(-r) = V(r)$. In other words, the real part of the complex potential must be a symmetric function of position, while the imaginary part should be antisymmetric. In fact a characteristic feature in PT-symmetric Hamiltonians is the reality of the eigenvalues below a threshold value of the potential (region where also eigenstates are PT-symmetric – usually called PT-phase), while above that threshold (in the so-called PT-broken phase) the PT-symmetry of the eigenstates is not preserved anymore and the eigenvalues become complex. At the transition point between PT and PT-broken phase, the so-called exceptional point (EP), two or more eigenvalues and the corresponding eigenvectors coincide, making EPs to be associated with exotic wave propagation behavior [5].

Although the PT-symmetry concept started from quantum mechanics, it was very quickly transferred to optics, where in recent years is being studied very actively [6-16]. In optical systems the PT-symmetry requirement results in symmetry conditions for the complex material parameters [6], i.e. permittivity and permeability, realizable by proper combination of loss and gain in a system. Since both gain and loss can be introduced and adjusted externally, it is much more straightforward to realize PT-symmetry in optical systems than in quantum ones, and this has been the main reason for the increasingly growing studies on optical PT-symmetric systems [6-27]. These studies revealed a series of novel and uncommon optical phenomena, such as unidirectional reflectionless perfect transmission [17], unidirectional invisibility [11, 15], PT-breaking transitions [19], simultaneous coherent perfect absorption (CPA) and lasing [18-20] and others.

Many of the above phenomena have been studied and demonstrated in scattering rather than paraxial propagation configuration. In such configurations the identification of the different PT-related phases is usually done through the eigenvalues of the scattering matrix [15, 17]. In the PT-symmetric phase the scattering matrix eigenvalues are unimodular, while in the PT-broken phase they are not unimodular but they form pairs of inverse moduli. At the EP two or more eigenvalues and eigenvectors coincide.

A major part of the existing studies regarding PT-symmetry in scattering configurations has been devoted also to the combination of PT-symmetry with metamaterials [21-25], and mainly with zero permittivity or permeability metamaterials [22, 23, 25], due to their unusual properties, such as constant phase throughout the whole metamaterial volume. Among the interesting relevant effects reported are nonreciprocal modes [25] at the interface between two mu-near-zero PT-symmetric metamaterials, exponentially bound interface modes characterized by zero attenuation in epsilon-near-zero metamaterials [23], etc.

A category of metamaterials that were until recently highly unexplored under the concept of PT-symmetry is chiral metamaterials [26-28], which are the subject of the present work. A chiral medium is such that it cannot be superimposed with its mirror image [29, 30]. This asymmetry results in a strong magneto-electric coupling, i.e. excitation of a magnetic polarization by an applied electric field and of an electric polarization by a magnetic field. As a result, the medium responds differently to left- and right-handed circularly polarized waves (LCP/RCP), which are the eigenwaves in chiral media. Among the interesting and useful effects resulting from this different response are the circular dichroism, i.e. different absorption of LCP and RCP waves, and the optical activity, i.e. the polarization rotation of an initially linearly polarized electromagnetic wave passing through the medium. (Both of these effects are due to the different refractive index (real and imaginary part) that LCP and RCP waves experience in chiral media). In chiral metamaterials the magneto-electric coupling is orders of magnitude stronger than in natural chiral media [30-36], because of their unit cell being much larger than atomic size; as a result, circular dichroism and optical activity are also much stronger in metamaterials, a feature which can be exploited in a variety of polarization control applications, such as polarization converters, polarization filters [35] etc.

Based on the above, by combining PT-symmetry (gain-loss and all the effects coming from their interplay, i.e. coherent perfect absorption-lasing, PT phase transition etc.) with chiral metamaterials (polarization related effects), two different worlds of propagation characteristics can be merged, thus promising unique propagation and polarization control possibilities. This paper aims to reveal part of these possibilities, some of which were also pointed out by our recent studies [26, 27].

More specifically, in this work we start by showing that in chiral media, despite the lack of mirror (and thus parity) symmetry, it is still possible to achieve full PT-symmetry and thus real eigenvalues; moreover we derive the necessary material parameter conditions for full PT-symmetry. Next we exploit these conditions in scattering configurations. In particular, we consider a simple model-system composed of a chiral PT-symmetric bilayer and we explore its scattering response under illumination with circularly polarized (CP) waves. First we investigate the case of normally incident waves. We find that the eigenvalues of the scattering matrix in this case are independent of chirality, i.e. the PT-symmetric phase is the same for chiral and non-chiral systems of the same permittivity and permeability values (obeying PT-symmetry). This is an important finding, as it allows us to tune the PT-related phenomena independently from the chirality-related phenomena. Next, we expand the analysis by allowing obliquely incident CP waves and we find that, now, the eigenvalues of the scattering matrix do depend on the chirality. Additionally, we find that any deviation from normal incidence opens up a rich phase diagram, including three distinct phases: full PT-symmetric phase, mixed PT-symmetric phase and broken PT-symmetric phase. As we demonstrate, both the boundaries of these phases and the polarization features of the waves transmitted through the system can be highly controlled by tuning the chirality parameter and/or the angle of incidence. Thus, as we demonstrate here, in PT-symmetric chiral metamaterials a wide range of advanced polarization control capabilities is possible.

## 2. Physical model and main equations

### 2.1. PT-symmetry conditions in chiral systems

In order to examine if there is a possibility to achieve PT-symmetry in chiral media and to derive the associated conditions we transform the Maxwell's curl equations, $\nabla \times \mathbf{E} = i\omega \mathbf{B}$ and $\nabla \times \mathbf{H} = -i\omega \mathbf{D}$ with the chiral constitutive relations [30, 37] $\mathbf{D} = \varepsilon\varepsilon_0 \mathbf{E} + i(\kappa/c)\mathbf{H}$ and $\mathbf{B} = \mu\mu_0 \mathbf{H} - i(\kappa/c)\mathbf{E}$ into an eigenvalue problem analogous to that of the Schrödinger's equation. (In the above relations $\varepsilon$, $\mu$, $\kappa$ refer to the relative permittivity, permeability and the chirality parameter respectively, $\varepsilon_0$, $\mu_0$ are the vacuum permittivity and permeability respectively and $c$ is the vacuum speed of light). Solving the resulting equations in terms of $\mathbf{E}$ and $\mathbf{H}$, the magneto-electric coupling in the constitutive relations does not allow for the problem to be cast as a simple eigenvalue problem, because the eigenvalue $\omega$ appears in the generalized Hamiltonian as well. A way to overcome the problem is to solve the equations for the fields $\mathbf{B}$, $\mathbf{D}$, instead. In that case we obtain an eigenvalue problem of the form

$$\hat{A} \begin{bmatrix} \mathbf{B} \\ \mathbf{D} \end{bmatrix} = \omega \begin{bmatrix} \mathbf{B} \\ \mathbf{D} \end{bmatrix} \tag{1}$$

where $\hat{A}$ is a tensor-operator analogous to the Hamiltonian in the Schrödinger problem, expressed as

$$\hat{A} = \begin{bmatrix} -i\nabla A_1(\mathbf{r}) \times -iA_1(\mathbf{r})\nabla \times & -i\nabla A_2(\mathbf{r}) \times -iA_2(\mathbf{r})\nabla \times \\ i\nabla A_3(\mathbf{r}) \times +iA_3(\mathbf{r})\nabla \times & i\nabla A_4(\mathbf{r}) \times +iA_4(\mathbf{r})\nabla \times \end{bmatrix} \tag{2}$$

In. Eq. (2) $A_1(\mathbf{r}) = -i\kappa c/(\varepsilon\mu - \kappa^2)$, $A_2(\mathbf{r}) = \mu\mu_0 c^2/(\varepsilon\mu - \kappa^2)$, $A_3(\mathbf{r}) = \varepsilon\varepsilon_0 c^2/(\varepsilon\mu - \kappa^2)$ and $A_4(\mathbf{r}) = i\kappa c/(\varepsilon\mu - \kappa^2)$ (for simplicity we have omitted the space dependence in $\varepsilon$, $\mu$ and $\kappa$). By requiring $\hat{A}$ to commute with the PT operator (see e.g. Ref. [21, 26]), i.e. $[\hat{P}\hat{T}, \hat{A}] = 0$, and taking into account that $\hat{P}\hat{T}(i\nabla \times) = i\nabla \times$, we derive the material parameter conditions as to have a PT-symmetric chiral system. These conditions read as follows:

$$\varepsilon(\mathbf{r}) = \varepsilon^*(-\mathbf{r}), \quad \mu(\mathbf{r}) = \mu^*(-\mathbf{r}), \quad \kappa(\mathbf{r}) = -\kappa^*(-\mathbf{r}) \tag{3}$$

Note that the first two of the conditions (3) are the ones that have been derived also in the case of non-chiral metamaterials [21].
   As has been already widely discussed though [5-19], the full PT-symmetry, i.e. the existence of real eigenvalues, requires that, besides the PT-symmetry of the Hamiltonian, $\hat{A}$, (ensured by the condition $[\hat{P}\hat{T},\hat{A}]=0$), also the eigenstates of $\hat{A}$ are PT-symmetric. In chiral media, where the eigenstates are left- and right-handed circularly polarized waves (denoted below by the subscript – and + respectively), the action of parity operator, $\hat{P}$, and time-reversal operator, $\hat{T}$, reveals that the eigenstates cannot be fully PT-symmetric. In particular, for eigenstates of the form $\mathbf{E}_\pm = E_0(\hat{x} \pm i\hat{y})e^{i(k_\pm z - \omega t)}$, where $E_0$ is the amplitude being a real number and $k_\pm = \omega(\sqrt{\varepsilon\mu} \pm \kappa)/c$ are the wavenumbers for RCP(+)/LCP(-) waves, the action of time-reversal operator, $\hat{T}$ ($i \to -i$, $t \to -t$), in $\mathbf{E}_\pm$ gives

$$\hat{T}\mathbf{E}_\pm = E_0(\hat{x} \mp i\hat{y})e^{-i(k_\pm^* z + \omega t)} \tag{4}$$

where the star (*) denotes complex conjugation. The action of parity operator, $\hat{P}$, in (4) flips the space $(x \rightarrow -x, y \rightarrow -y, z \rightarrow -z)$, i.e.

$$\hat{P}\hat{T}\boldsymbol{E}_\pm = -E_0(\hat{x} \mp i\hat{y})e^{i(k_\pm^* z - \omega t)} \qquad (5)$$

Employing the conditions (3) to Eq. (5), to associate the wavenumbers $k_\pm$ with their PT-symmetric counterparts, we conclude that

$$\hat{P}\hat{T}\boldsymbol{E}_\pm = -\boldsymbol{E}_\mp. \qquad (6)$$

Thus, the action of $\hat{P}\hat{T}$ in our case does not leave the eigenwaves unchanged; it rather transforms an initial RCP wave to LCP and vice versa - except for a phase factor. (Note that using the conditions (3) the above PT transformations lead to the same conclusion also for the fields *H*, *B* and *D*.) Nevertheless, despite the lack of PT-symmetry in the eigenwaves, there is still the possibility for real eigenvalues (PT-symmetric phase). This possibility is ensured by the degeneracy of RCP/LCP eigenwaves with respect to the frequency, as can be easily seen by applying the condition (3) into Eq. (1) [26]. Hence, our analysis demonstrates the possibility to obtain real eigenvalues in PT-symmetric chiral systems and provides the associated conditions. However, as numerical results in scattering systems have shown, the conditions (3) for the chirality parameter are not the only ones that lead to fully PT-symmetric phase, at least in scattering configurations [26]. There, there are cases beyond condition (3) for $\kappa$ which yield similar features and similarly useful properties [27].

*2.2. Chiral systems in scattering configurations under normal incidence: The scattering matrix*

Having derived the necessary conditions for PT-symmetry in chiral systems [Eq. (3)], in the rest of the paper we turn our attention to scattering configurations, applying those conditions and investigating the scattering features and response of the resulting systems. Our model-system is a homogeneous one-dimensional (1D) chiral bilayer (of thickness $L = 2d$ along $z$ and infinite extent along the $x, y$ directions), as shown in Fig. 1. The complex material parameters $\varepsilon$, $\mu$, and $\kappa$ of the two media involved in our system are the relative permittivity, relative permeability and chirality parameter, respectively.

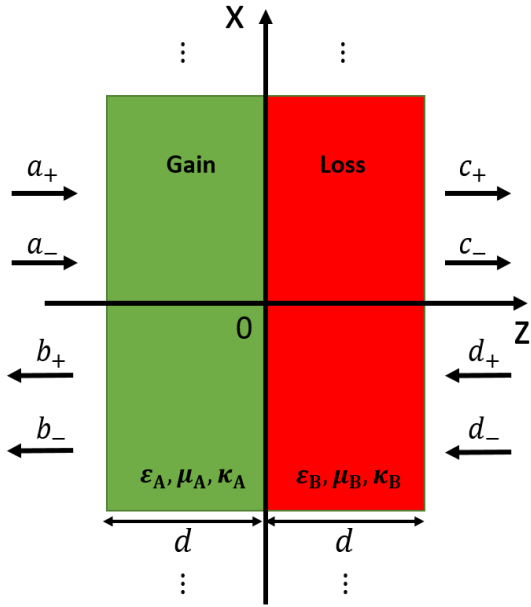

**Figure 1.** Our model one-dimensional chiral bilayer. It is infinite along the *x*- and *y*- directions and of thickness *L*=2*d* along the *z*-direction. For PT-symmetry the material parameters of the two media (A and B) should obey $\varepsilon_A = \varepsilon_B^*$, $\mu_A = \mu_B^*$, $\kappa_A = -\kappa_B^*$, with $\varepsilon$, $\mu$, $\kappa$ being the relative permittivity, permeability and chirality parameter. The amplitudes of the incident ($a_\pm, d_\pm$) and scattered ($b_\pm, c_\pm$) waves are also shown, where the subscripts +/- account for RCP/LCP waves.

We consider in our system normal incidence with four input and four output ports, each one allowing either left circularly polarized (LCP) or right circularly polarized (RCP) waves as illustrated in Fig. 1. We start with the most general form for the electric field outside the PT-symmetric structure, as

$$E(z,t) = \begin{cases} (a_+ e^{ik_0 z} + b_+ e^{-ik_0 z})\hat{e}_+ + (a_- e^{ik_0 z} + b_- e^{-ik_0 z})\hat{e}_-, & z < -d \\ (c_+ e^{ik_0 z} + d_+ e^{-ik_0 z})\hat{e}_+ + (c_- e^{ik_0 z} + d_- e^{-ik_0 z})\hat{e}_-, & z > d \end{cases} \quad (7)$$

where $a_\pm, d_\pm$ and $b_\pm, c_\pm$ are amplitudes of the ingoing and outgoing, respectively, LCP (-) and RCP (+) waves (as observed from the source point), $\hat{e}_+ = (\hat{x} + i\hat{y})/\sqrt{2}$ and $\hat{e}_- = (\hat{x} - i\hat{y})/\sqrt{2}$, and $k_0 = \omega/c$. Similar expression can be used for the electric field inside the chiral bilayer, with $k_0$ replaced by the relevant chiral wavenumbers $k_\pm = \omega(\sqrt{\varepsilon\mu} \pm \kappa)/c$ [30].

Due to the two possible circular polarizations at each of the two sides of the structure, the system should be described by a 4x4 scattering matrix $S_0$, defined by (see Fig. 1)

$$\begin{pmatrix} b_- \\ c_+ \\ b_+ \\ c_- \end{pmatrix} = S_0 \begin{pmatrix} a_+ \\ d_- \\ a_- \\ d_+ \end{pmatrix} \equiv \begin{pmatrix} r_{-+}^{(L)} & t_{--}^{(R)} & r_{--}^{(L)} & t_{-+}^{(R)} \\ t_{++}^{(L)} & r_{+-}^{(R)} & t_{+-}^{(L)} & r_{++}^{(R)} \\ r_{++}^{(L)} & t_{+-}^{(R)} & r_{+-}^{(L)} & t_{++}^{(R)} \\ t_{-+}^{(L)} & r_{--}^{(R)} & t_{--}^{(L)} & r_{-+}^{(R)} \end{pmatrix} \begin{pmatrix} a_+ \\ d_- \\ a_- \\ d_+ \end{pmatrix} = \begin{pmatrix} r^{(L)} & t_{--} & 0 & 0 \\ t_{++} & r^{(R)} & 0 & 0 \\ 0 & 0 & r^{(L)} & t_{++} \\ 0 & 0 & t_{--} & r^{(R)} \end{pmatrix} \begin{pmatrix} a_+ \\ d_- \\ a_- \\ d_+ \end{pmatrix} \quad (8)$$

where the last equation shows the $S_0$ matrix as is simplified in our case. In Eq. (8) $t_{++} \equiv t_{++}^{(L)} = t_{++}^{(R)}$ and $t_{--} \equiv t_{--}^{(L)} = t_{--}^{(R)}$ are the transmission coefficients for RCP(+)/LCP(-) light while $r^{(L)} \equiv r_{+-}^{(L)} = r_{-+}^{(L)}$ and $r^{(R)} \equiv r_{+-}^{(R)} = r_{-+}^{(R)}$ are the reflection coefficients for LCP(-)/RCP (+) light (the superscript (L) or (R) indicates incidence from left or right, respectively). The remaining eight scattering coefficients ($t_{+-}^{(L)}, t_{-+}^{(L)}, t_{+-}^{(R)}, t_{-+}^{(R)}$ and $r_{++}^{(L)}, r_{--}^{(L)}, r_{++}^{(R)}, r_{--}^{(R)}$) are zero in our case. (Note that the first subscript in $t$ and $r$ indicates the scattered wave polarization and the second the incident wave polarization.)

To investigate when the system lies in PT-symmetric or PT-broken phase and to identify the position of exceptional points the first step is to evaluate the eigenvalues and eigenvectors of the scattering matrix $S_0$ [17, 38, 39]. As was mentioned also in the Introduction, in the PT-symmetric phase the eigenvalues of the scattering matrix are unimodular, while in the PT-broken phase they are not. Therefore by calculating the scattering matrix eigenvalues one can identify the different possible PT-related phases and the EP. The eigenproblem solution for our scattering matrix $S_0$ yields two degenerate pairs of eigenvalues, $\sigma$, which are given by

$$\sigma_{1,2} = \sigma_{3,4} = \frac{1}{2}\left(r^{(L)} + r^{(R)} \pm \sqrt{(r^{(L)} - r^{(R)})^2 + 4t_{++}t_{--}}\right). \quad (9)$$

The corresponding eigenvectors are given by

For $\sigma_{1,2}$: $v_{1,2} = \begin{pmatrix} \frac{1}{2t_{--}}\left(r^{(L)} - r^{(R)} \pm \sqrt{(r^{(L)} - r^{(R)})^2 + 4t_{++}t_{--}}\right) \\ 1 \\ 0 \\ 0 \end{pmatrix}$ (10a)

for $\sigma_{3,4}$: $v_{3,4} = \begin{pmatrix} 0 \\ 0 \\ \frac{1}{2t_{++}}\left(r^{(L)} - r^{(R)} \pm \sqrt{(r^{(L)} - r^{(R)})^2 + 4t_{++}t_{--}}\right) \\ 1 \end{pmatrix}$ (10b)

From Eqs. (8), (10) and Fig. 1 one can observe that the eigenvectors correspond to pure circularly polarized waves (either incoming or outgoing) of opposite handedness at the two opposite sides of the slab.

Note that depending on the arrangement of the input and output ports in the column matrices of Eq. (8), we can build our scattering matrix formalism in several ways, which characterize different physical processes. Here, we have chosen the case related with the position of exceptional points (EPs), in which the system passes from PT-symmetric phase to PT-broken phase [17-19, 39] (similarly to non-chiral PT symmetric systems) [19]. In this case, the scattering matrix $S_0$ satisfies the relation $PTS_0(\omega^*)PT = S_0^{-1}(\omega)$ [17, 19], which is a fundamental condition obeyed by $PT$-symmetry.

Although for PT-symmetry we should impose conditions (3) to the material parameters of our system, to obtain more general expressions we start our analysis with arbitrary parameters $\varepsilon_A, \mu_A, \kappa_A, \varepsilon_B, \mu_B, \kappa_B$. For arbitrary material parameters the reflection and transmission coefficients for circularly polarized waves are expressed as

$$t_{++} = -\frac{8e^{id(k_A^+ + k_B^+ - 2k_0)}}{A+B} \tag{11a}$$

$$t_{--} = -\frac{8e^{id(k_A^- + k_B^- - 2k_0)}}{A+B} \tag{11b}$$

$$r^{(L)} = -\frac{e^{-2ik_0 d}(C^{(L)} + D^{(L)})}{A+B} \tag{11c}$$

$$r^{(R)} = -\frac{e^{-2ik_0 d}(C^{(R)} + D^{(R)})}{A+B} \tag{11d}$$

where

$$A = Z_A \left[ +\left(\frac{1}{Z_A}+1\right) + \left(\frac{1}{Z_A}-1\right) e^{id(k_A^+ + k_A^-)} \right] \left[ -\left(\frac{1}{Z_B}+1\right) + \left(\frac{1}{Z_B}-1\right) e^{id(k_B^+ + k_B^-)} \right] \tag{12a}$$

$$B = Z_B \left[ -\left(\frac{1}{Z_A}+1\right) + \left(\frac{1}{Z_A}-1\right) e^{id(k_A^+ + k_A^-)} \right] \left[ +\left(\frac{1}{Z_B}+1\right) + \left(\frac{1}{Z_B}-1\right) e^{id(k_B^+ + k_B^-)} \right] \tag{12b}$$

$$C^{(L)} = Z_A \left[ +\left(\frac{1}{Z_A}+1\right) e^{id(k_A^+ + k_A^-)} + \left(\frac{1}{Z_A}-1\right) \right] \left[ +\left(\frac{1}{Z_B}-1\right) e^{id(k_B^+ + k_B^-)} - \left(\frac{1}{Z_B}+1\right) \right] \tag{12c}$$

$$D^{(L)} = Z_B \left[ +\left(\frac{1}{Z_A}+1\right) e^{id(k_A^+ + k_A^-)} - \left(\frac{1}{Z_A}-1\right) \right] \left[ +\left(\frac{1}{Z_B}-1\right) e^{id(k_B^+ + k_B^-)} + \left(\frac{1}{Z_B}+1\right) \right] \tag{12d}$$

$$C^{(R)} = Z_A \left[ +\left(\frac{1}{Z_B}+1\right) e^{id(k_B^+ + k_B^-)} - \left(\frac{1}{Z_B}-1\right) \right] \left[ +\left(\frac{1}{Z_A}-1\right) e^{id(k_A^+ + k_A^-)} + \left(\frac{1}{Z_A}+1\right) \right] \tag{12e}$$

$$D^{(R)} = Z_B \left[ +\left(\frac{1}{Z_B}+1\right) e^{id(k_B^+ + k_B^-)} + \left(\frac{1}{Z_B}-1\right) \right] \left[ +\left(\frac{1}{Z_A}-1\right) e^{id(k_A^+ + k_A^-)} - \left(\frac{1}{Z_A}+1\right) \right] \tag{12f}$$

and $k_i^\pm = k_0(\sqrt{\varepsilon_i \mu_i} \pm \kappa_i)$, $i = A$ or $B$, are the wavevectors while $Z_i = \sqrt{\mu_0 \mu_i / \varepsilon_0 \varepsilon_i}$, $i = A$ or $B$, are the wave impedances in the two chiral layers.

Investigating further the above equations, one can observe that the *reflection coefficients are totally independent of the chirality parameters* of the two media, since all the terms $A, B, C^{(L)}, D^{(L)}, C^{(R)}, D^{(R)}$ are chirality independent (note that they depend only on the impedances $Z_A, Z_B$ and terms of the form $e^{id(k_A^+ + k_A^-)} = e^{2idk_0\sqrt{\varepsilon_A \mu_A}}$ and $e^{id(k_B^+ + k_B^-)} = e^{2idk_0\sqrt{\varepsilon_B \mu_B}}$, all independent of chirality). Moreover, the product $t_{++}t_{--}$ is also independent of chirality. (Note that $t_{++} = t_{non-chiral} e^{idk_0(\kappa_A + \kappa_B)}$ and $t_{--} = t_{non-chiral} e^{-idk_0(\kappa_A + \kappa_B)}$, where $t_{non-chiral}$ is the transmission for the system of the same permittivities and permeabilties but with zero chirality for both media.) Therefore the *eigenvalues of the scattering matrix* are *independent of chirality*, demonstrating that the different PT-symmetry-related phases of our system and the exceptional point are totally chirality independent. This is an important feature, which will be demonstrated and discussed further in the next sections.

*2.3. Generalized unitarity relations for chiral PT-symmetric systems*

By exploiting the PT-symmetry properties of our system following the procedures of [17], we can formulate the generalized unitarity relation for the four-port case. This can be done using the transfer matrix formalism, which is commonly used in electromagnetic theory [17, 18]. We start from the general 4x4 scattering matrix configuration (Eq. 8), expressing the elements of the transfer matrix $M$ in terms of the elements of the scattering matrix $S_0$ as

$$M = \begin{pmatrix} 1/S_{21} & -S_{22}/S_{21} & 0 & 0 \\ S_{11}/S_{21} & S_{12} - (S_{11}S_{22}/S_{21}) & 0 & 0 \\ 0 & 0 & 1/S_{43} & -S_{44}/S_{43} \\ 0 & 0 & S_{33}/S_{43} & S_{34} - (S_{33}S_{44}/S_{43}) \end{pmatrix} \quad (13)$$

Therefore, the transfer matrix $M$ can be written as

$$\begin{pmatrix} a_+ \\ b_- \\ a_- \\ b_+ \end{pmatrix} = M \begin{pmatrix} c_+ \\ d_- \\ c_- \\ d_+ \end{pmatrix} = \begin{pmatrix} 1/t_{++} & -r^{(R)}/t_{++} & 0 & 0 \\ r^{(L)}/t_{++} & t_{--} - (r^{(L)}r^{(R)}/t_{++}) & 0 & 0 \\ 0 & 0 & 1/t_{--} & -r^{(R)}/t_{--} \\ 0 & 0 & r^{(L)}/t_{--} & t_{++} - (r^{(L)}r^{(R)}/t_{--}) \end{pmatrix} \begin{pmatrix} c_+ \\ d_- \\ c_- \\ d_+ \end{pmatrix} \quad (14)$$

It is straightforward to show that the transfer matrix $M$ satisfies the quite general condition $\det[M] = 1$ [17, 18] and, additionally, the symmetry relation $M = (M^{-1})^*$ which is a direct consequence of PT-symmetry. By taking into account the above two conditions, resulting to $MM^* = I$ (with I the identity 4x4 matrix), we obtain the *generalized unitarity relation* for a general chiral PT-symmetric system, which is expressed as

$$\left|\sqrt{T_{++}T_{--}} - 1\right| = \sqrt{R^{(L)}R^{(R)}} \quad (15)$$

where $T_{++} \equiv |t_{++}|^2$, $T_{--} \equiv |t_{--}|^2$ and $R^{(L)} \equiv |r^{(L)}|^2$, $R^{(R)} \equiv |r^{(R)}|^2$ are the transmission and reflection power coefficients, respectively. It should be noted that the relation (15) is valid for all 1D chiral PT-symmetric structures and it holds in both PT-symmetric and PT-broken phase. It does not correspond though to flux conservation, which can occur only under strict conditions regarding the incident wave configurations.

Since, as was pointed out earlier, both the reflection coefficients and the product $t_{++}t_{--}$ are independent of chirality, all the terms of Eq. (15) are chirality-independent, although the coefficients $T_{++}$ and $T_{--}$ alone are chirality-dependent. This shows that if we add chirality in a non-chiral PT-symmetric bilayer which is at an anisotropic transmission resonance (ATR) point (i.e. $T = 1$ and one of the $R^{(L)}$, $R^{(R)}$ is zero), which corresponds to unidirectional flux conservation, we will always be at the condition $T_{++}=1/T_{--}$. In that case, though, a flux-conserving process for waves incident from one side is realized only if $T_{++}=T_{--} = 1$, a condition incompatible with the PT-symmetry requirement for $\kappa$ (for non-zero Im($\kappa$)) and normal incidence [see Eqs. (11) and paragraph after them].

**3. Results and Discussion**

*3.1. PT-symmetric chiral bilayer under normal incidence: Different PT phases and scattering characteristics*

In non-Hermitian systems the amplitude of the scattering matrix eigenvalues is either below (decay) or above (growth) unity. In the case of PT-symmetry, where gain and loss are in balance, the eigenvalues of the scattering matrix can be unimodular and are in fact unimodular below the exceptional point; thus their calculation offers the possibility to identify the exceptional point and the different PT-related phases. To investigate this possibility and its associated effects in the case of chiral systems, we investigated the system of Fig. 1 with some representative material parameters satisfying the PT-symmetry conditions [Eq. (3)]: $\varepsilon(z) = 3.0 + 0.4i$, $\varepsilon(-z) = 3.0 - 0.4i$, $\mu(z) = 1.1 + 0.1i$, $\mu(-z) = 1.1 - 0.1i$ and $\kappa(z) = 0.04 - 0.04i$, $\kappa(-z) = -0.04 - 0.04i$ for $-d<z<d$. For those parameters the scattering matrix ($S_0$) eigenvalues (see Eq. (9)) as a function of frequency are plotted in Fig. 2(e), demonstrating the existence of the two different phases, the PT-symmetric one characterized by unimodular eigenvalues and the PT-broken one, characterized by eigenvalues of inverse moduli. At the exceptional point (at $\omega L/c = 22.25$) all eigenvalues coincide.

As has been discussed already in Section 2, the eigenvalues in our case are independent of chirality, since both the reflection coefficients and the transmission product $t_{++}t_{--}$ are chirality independent. This is also verified numerically in Fig. 2, where besides the eigenvalues of the chiral PT system we have also plotted the eigenvalues of a system with the above PT-symmetric permittivity and permeability values but with $\kappa=0$ in both media (see Fig. 2(a)) and for a system without any symmetry in $\kappa$, i.e. $\kappa(-z) = -0.06 - 0.02i$, $\kappa(z) = -0.04 - 0.01i$ (see Fig. 2(i)). For all three systems the scattering matrix eigenvalues shown in Fig. 2 are identical.

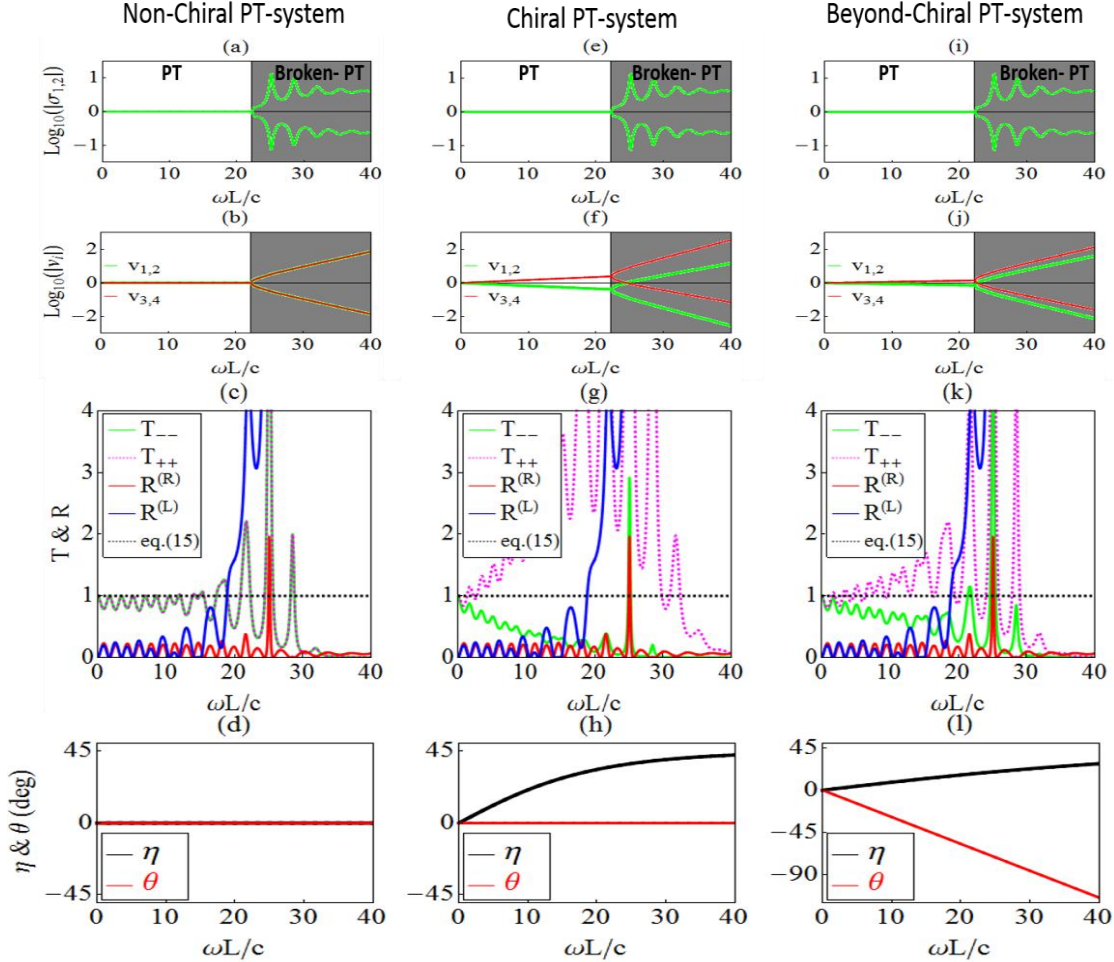

**Figure 2.** Left column: PT-symmetric bilayer (see Fig. 1) of length $L$ without chirality. Middle column: chiral PT-symmetric bilayer. Right column: chiral bilayer beyond PT-symmetry (for the material parameters of the systems see text). First row: Eigenvalues of the scattering matrix $S_0$ (see Eq. (9)). Second row: Eigenvectors of the scattering matrix $S_0$ (see Eq. (10)). Third row: Reflection, $R$, and transmission, $T$, (power) coefficients for LCP(-) and RCP(+) waves impinging on both left (see superscript (L)) and right (see superscript (R)) sides of the systems. The generalized unitary relation, Eq. (15), is also plotted, as $R^{(L)}R^{(R)} + 2\sqrt{T_{++}T_{--}} - T_{++}T_{--} = 1$ (dashed line). Bottom row: Ellipticity, $\eta$, and optical activity, $\theta$, of a wave transmitted through the three systems studied.

All features are plotted as a function of the dimensionless frequency $\omega L/c$, with $c$ the vacuum speed of light and $L=2d$ the total system length/thickness.

For the above-mentioned three systems we have calculated also the amplitudes of the scattering matrix eigenvectors (second row of Fig. 2) and the transmission ($T$) and reflection ($R$) power coefficients for circularly polarized incident waves (third row of Fig. 2). From the eigenvector plots (depicting the non-zero, non-unity components of Eqs. (10)) it can be observed that while in the case of non-chiral PT-symmetric systems (Fig. 2(b)) the ratio of the two non-vanishing components for each eigenvector is unimodular below the exceptional point, in chiral PT-symmetric structures (Fig. 2(f)) it is not. Considering a system excitation configuration of the form of an eigenvector, the corresponding scattered waves, which are pure circularly polarized waves of opposite handedness at the left and right side of the slab, above the exceptional point will be either exponentially growing or attenuating (depending on the incident wave configuration). At particular frequencies above EP these two modes (growing-attenuating) are expected to give simultaneous *CPA and lasing of circularly polarized waves*, analogous with the CPA-lasing modes of the non-chiral PT-symmetric systems [18-19]. The particular circular polarization favored in our foreseen CPA-laser modes depends mainly on the sign of the Im($\kappa$) of the two chiral slabs. We have to note here that the above described eigenvector behavior seem to survive qualitatively even if the chiral system is beyond PT-symmetry (see Fig. 2(j)).

Regarding the transmission and reflection coefficients, for our isotropic chiral medium, as expected, $T_{++}$ and $T_{--}$ do not depend on the side of incidence (since the system is reciprocal), while a similar condition does not hold for the reflection coefficients, as can be seen in Fig. 2(g). Comparing Fig. 2(g) with Figs. 2(c) and 2(k) (i.e. the case of non-chiral PT-symmetric structure and the case beyond PT-symmetry in $\kappa$, respectively), we can see that the chirality parameter has strong influence on the transmission coefficient and none on the reflection. Note that the chirality independence of the reflection coefficient was also observed in the corresponding analytical expressions (even for non-PT systems - see Section 2), while from the same expressions (see Eqs. (11), (12) and paragraph after them) one can conclude that the transmission (power) coefficient depends on the chirality parameter exponentially (in particular $T$ depends on the Im($\kappa_A+\kappa_B$) – see Fig. 1). Such a dependence, although having strong influence on the transmission magnitude, does not affect the frequency position of the zeros and the resonances of the transmission, something observable also from Fig. 2.

In the rest of the current subsection we investigate the impact of our system on the polarization of an incoming wave. For that, we examine first the optical activity, $\theta$, of our structure, which for absence of cross-polarized transmission terms (e.g. $t_{+-}$) is given by $\theta = 0.5[\text{Arg}(t_{++}) - \text{Arg}(t_{--})]$. Taking into account Eqs. (11a) and (11b), we can find that for a general chiral bilayer as the one of Fig. 1 (i.e. prior application of PT-symmetry conditions)

$$\theta = k_0 d[\text{Re}(\kappa_A) + \text{Re}(\kappa_B)] \tag{16}$$

Due to the symmetries imposed by applying PT [Eq. (3)], the real part of the chirality $\kappa$ changes sign across the bilayer. Hence, the polarization rotation occurring in the gain slab is subsequently canceled out when the wave passes through the loss slab, resulting to zero optical rotation, as is demonstrated also in Fig. 2(h). Next we consider the other important polarization-related property, which is the circular dichroism (CD). A quantity directly related with the CD response is the transmitted wave ellipticty, $\eta$, for a linearly polarized incident wave, which is given by

$$\eta = 0.5\sin^{-1}\left[\frac{(T_{++}-T_{--})}{(T_{++}+T_{--})}\right] = 0.5\sin^{-1}\left[\frac{1-\exp[4k_0 d(\text{Im}(\kappa_A)+\text{Im}(\kappa_B))]}{1+\exp[4k_0 d(\text{Im}(\kappa_A)+\text{Im}(\kappa_B))]}\right] \tag{17}$$

(The second r.h.s. of Eq. (17) has been obtained employing equations (11), concerning general material parameters in the two slabs). In a chiral PT-symmetric system, the imaginary part of chirality preserves its sign across the system implying a very strong circular dichroism (CD), resulting in a large degree of the transmitted wave ellipticity, as shown/confirmed also in Fig. 2(h).

Summarizing, as shown in Fig. 2 and discussed also in connection with analytical calculations, the PT-symmetric features of a chiral PT-symmetric bilayer (i.e. different PT-phases, exceptional point, structure resonances) are totally independent of chirality. On the other hand, the chirality strongly affects the polarization state of the wave passing through the bilayer (through its effect to its ellipticity). This shows that if one has the ability to control separately the system permittivity/permeability and the system chirality (as is to a large extent possible in chiral metamaterials) one can combine or superimpose almost at will the PT-related and the chirality-related features, achieving fascinating or important in applications effects, e.g. CPA-lasing for circularly polarized waves.

*3.2. PT-symmetric chiral bilayer under oblique incidence: Controlling the PT-symmetry phase*

As has been already discussed, in the case of waves normally incident on the structure of Fig. 1 the position of EP, which characterizes the transition from the PT-symmetric to the PT-broken phase, is totally independent of chirality. As we discuss in this section, the situation changes for oblique incidence, where chirality strongly affects the different PT-related phases and the EPs.

Since, as was shown recently [40], transverse electric (TE) and transverse magnetic (TM) polarized waves are associated with different exceptional points, it follows as a result that for circularly polarized light under oblique incidence a mixed phase is realizable. Therefore, it is interesting to examine the different PT-related phases and the phase transitions for CP light obliquely incident on our chiral bilayer structure. In this case, the transmission and reflection coefficients $t_{+-}^{(L)}, t_{-+}^{(L)}, r_{++}^{(L)}, r_{--}^{(L)}$ and $t_{+-}^{(R)}, t_{-+}^{(R)}, r_{++}^{(R)}, r_{--}^{(R)}$ are not zero anymore, and, hence the most general scattering matrix, $S_0$ (see Eq. (8)), should be considered [27]. By numerically calculating the eigenvalues of this scattering matrix the attainable PT-related phases of the bilayer can be identified.

As an example, we investigate here a system with permittivity and permeability values the same as in Fig. 2, i.e. $\varepsilon(z) = 3.0 + 0.4i$, $\varepsilon(-z) = 3.0 - 0.4i$, $\mu(z) = 1.1 + 0.1i$, $\mu(-z) = 1.1 - 0.1i$. In Fig. 3 we plot the eigenvalues of the scattering matrix $S_0$ as a function of the normalized frequency, $\frac{\omega L}{c}$, at incidence angle $\theta_{in} = 45°$ and three different cases regarding chirality: (a) without chirality $\kappa(z) = \kappa(-z) = 0$, (b) with chirality that respects PT-symmetry, i.e. $\kappa(z) = 0.04 - 0.04i$, $\kappa(-z) = -0.04 - 0.04i$ and (c) with $\kappa(z) = -0.04 - 0.01i$, $\kappa(-z) = -0.06 - 0.02i$, i.e., beyond PT-symmetry. It can be observed that a consequence of obliquely incident waves is the appearance of mixed phases where one pair of eigenvalues is unimodular while the other is not. Moreover, in contrast to what happens for normal incidence, for oblique incidence the positions of the EPs (two in this case) are strongly affected by chirality. *This implies that one important impact of chirality on the PT-related features of our structure is the tuning of the EPs.*

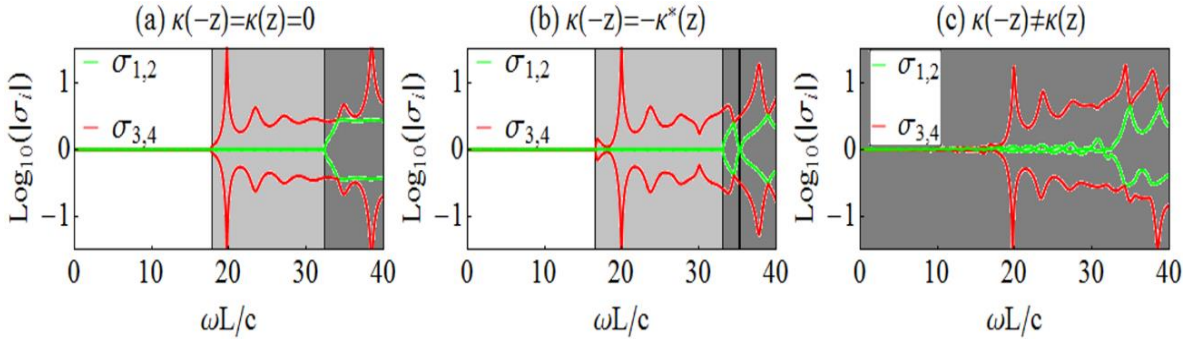

**Figure 3.** Eigenvalues ($\sigma$) of the scattering matrix $S_0$ for the chiral bilayer shown in Fig. 1, for incidence angle $\theta_{in} = 45°$, as a function of the dimensionless frequency $\omega L/c$, where $L$ is the system length and $c$ the vacuum speed of light. The permittivity and permeability of the two media are as in Fig. 2. Panel (a): Simple PT-symmetric system without chirality. Panel (b): Chiral PT-symmetric system. Panel (c): System beyond PT-symmetry, $\kappa(-z) \neq -\kappa^*(z)$.

To analyze further the impact of chirality on the different PT–related phases of a chiral bilayer we investigate the different PT-related phases for a system with permittivities and permeabilities as in Fig.

3 by scanning the PT-obeying chirality parameter (both real and imaginary part) at a fixed frequency, $\omega L/c = 15.5$, and incidence angle $\theta_{in} = 45°$. The results for the eigenvalues of the scattering matrix as a function of chirality are illustrated in Fig. 4(a). In Fig. 4(a) we can see that as we increase the chirality the system passes from PT-symmetric phase to mixed PT-symmetric phase (light grey) and, with further increase of the chirality, to PT-broken phase, indicating a possibility of a full control of exceptional points (EPs) and the associated PT phases by changing the chirality. Surprisingly, it is possible to achieve analogous control by changing only the real or only the imaginary part of the chirality, as illustrated in Figs. 4(b) and 4(c), respectively. All the above reveal a very rich behavior and possibility for different phases and phase re-entries as one changes the system chirality.

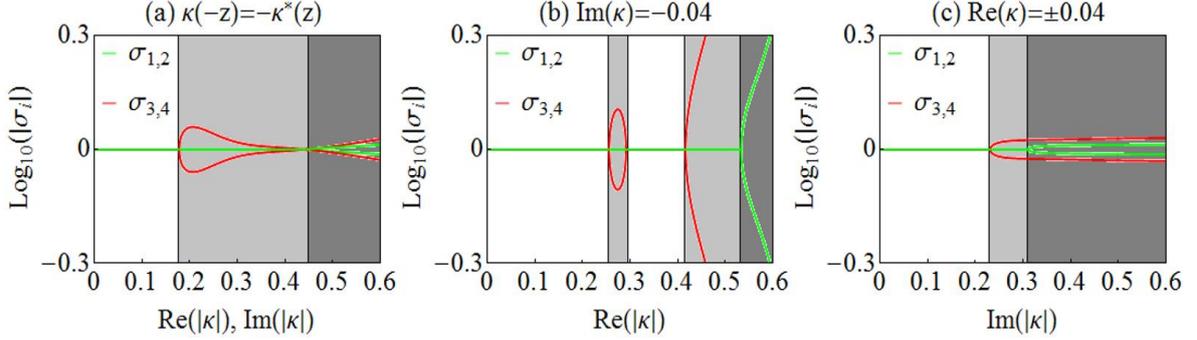

**Figure 4.** Eigenvalues ($\sigma$) of the scattering matrix $S_0$ at incidence angle $\theta_{in} = 45°$ as a function of the chirality parameter, $\kappa$, for chiral PT-symmetric systems as the one shown in Fig. 1 and for dimensionless frequency $\omega L/c = 15.5$. The permittivity and permeability of the two media are as in Fig. 2. Panel (a): Scan of the chirality parameter both real and imaginary parts; Panel (b): Scan of the real part of the chirality for $\text{Im}(\kappa) = -0.04$. Panel (c): Scan of the imaginary part of the chirality for $\text{Re}(\kappa) = \pm 0.04$.

Although for oblique incidence the PT-related features of a chiral bilayer can be highly controlled by chirality, however chirality can not offer an external dynamic control. Such a control can be offered by the angle of incidence as we show in Fig. 5. In Fig. 5 we plot the phase diagrams of the same systems as in Figs. 3(a) (see panels (a), (b) of Fig. 5) and 3(b) (see panels (c) and (d)) as we change the incidence angle. As can be observed there, the frequency positions of the exceptional points and, consequently, the frequency extent of the different PT-phases are highly dependent on the incidence angle. Moreover, incident angle controllable mixed phases and phase re-entries (see panel (d)) can be achieved. The possibility of PT-feature tuning by the incidence angle offers a great and practical way for a dynamic control of chiral PT-symmetric systems.

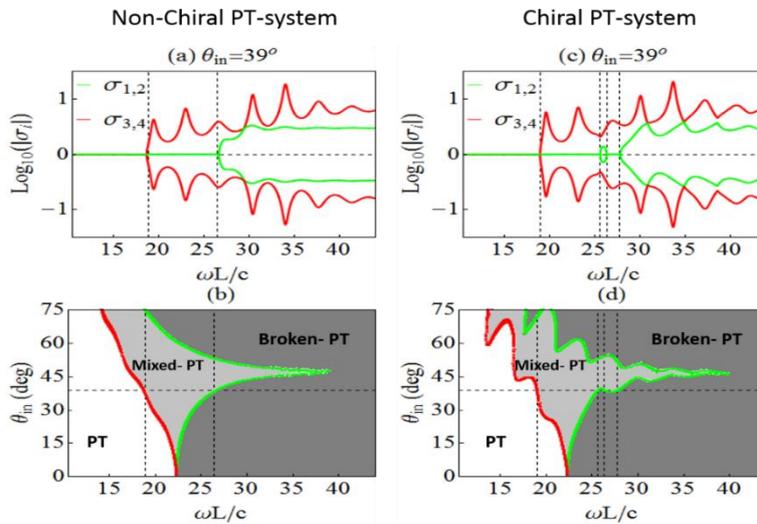

**Figure 5.** PT-symmetric system as the one of Fig. 1 (of length $L$) without chirality (left column) and with chirality (right column), with permittivity and permeability as in Fig. 2. (a) & (c): Eigenvalues ($\sigma$) of the scattering

matrix $S_0$ at incidence angle $\theta_{in} = 39^o$, as a function of the dimensionless frequency $\omega L/c$, with $c$ the vacuum speed of light. (b) & (d): Phase diagrams showing the different attainable phases for different incidence angles. The vertical dashed lines corresponds to the exceptional points of plots (a) and (c).

As was mentioned already, among the interesting and particularly useful characteristics associated with chiral media are the optical activity and the circular dichroism (CD), which for our double-layer chiral slab and for normal incidence depend exclusively on the chirality parameter of the two layers (see Eqs. (16), 17)). In the case of oblique incidence, where in the chiral bilayer there are cross-polarized transmission terms even for circularly polarized incident waves, the optical activity and the transmitted wave ellipticity (a measure of the CD response) cannot be obtained anymore by the simple relations (16) and (17). They can be calculated though through the Stokes parameters [27,29], which describe completely the polarization state of a wave. The four Stokes parameters for the transmitted wave in our case are defined by $S_0 = E_\perp E_\perp^* + E_\| E_\|^*$, $S_1 = E_\perp E_\perp^* - E_\| E_\|^*$, $S_2 = 2\text{Re}[E_\perp E_\|^*]$ and $S_3 = 2\text{Im}[E_\perp E_\|^*]$, where the star (*) denotes complex conjugation, the subscript ∥ indicates the transmitted electric field component that lies on the plane of incidence while the subscript ⊥ indicates the perpendicular component. Through Stokes parameters the optical activity is given by

$$\theta = \frac{1}{2}\tan^{-1}\left(\frac{S_2}{S_1}\right), \quad (0 \leq \theta \leq \pi) \tag{18}$$

and the ellipticity by

$$\eta = \tan^{-1}\left(\frac{S_3/S_0}{1+[1-(S_3/S_0)^2]^{1/2}}\right), \quad (-\pi/4 \leq \eta \leq \pi/4) \tag{19}$$

To be able to calculate $\theta$ and $\eta$ employing the above formulas and to have a full picture for our system potential for polarization manipulation, one needs to calculate the scattering coefficients (transmission and reflection) for linearly polarized incident waves. These coefficients can be directly obtained from the circularly polarized reflection and transmission data according to the equations [30]

$$\begin{pmatrix} t_{\|\|} & t_{\|\perp} \\ t_{\perp\|} & t_{\perp\perp} \end{pmatrix} = \frac{1}{2}\begin{pmatrix} (t_{++} + t_{--}) + (t_{+-} + t_{-+}) & -i[(t_{++} - t_{--}) - (t_{+-} - t_{-+})] \\ i[(t_{++} - t_{--}) + (t_{+-} - t_{-+})] & (t_{++} + t_{--}) - (t_{+-} + t_{-+}) \end{pmatrix} \tag{20}$$

and

$$\begin{pmatrix} r_{\|\|} & r_{\|\perp} \\ r_{\perp\|} & r_{\perp\perp} \end{pmatrix} = \frac{1}{2}\begin{pmatrix} (r_{+-} + r_{-+}) + (r_{++} + r_{--}) & -i[(r_{+-} - r_{-+}) - (r_{++} - r_{--})] \\ i[(r_{+-} - r_{-+}) + (r_{++} - r_{--})] & (r_{+-} + r_{-+}) - (r_{++} + r_{--}) \end{pmatrix} \tag{21}$$

In Eqs. (20), (21), as in the Stokes parameters, the subscripts ∥ and ⊥ indicate the components parallel and perpendicular to the plane of incidence, respectively, while, as in the circular polarization case, the first subscript refers to the transmitted (or reflected) component and the second to the incident one.

In Fig. 6 we show the transmitted and reflected power coefficients ($T=|t|^2$, $R=|r|^2$ respectively) for linearly polarized waves as well as the corresponding optical activity and ellipticity for the chiral PT-symmetric system examined in Fig. 2 under oblique incidence. In particular, we present those quantities at $\omega L/c = 15.5$ as a function of chirality (left two columns) and as a function of the incidence angle (right two columns) keeping all the other parameters constant. Our results reveal a very rich behavior of propagation characteristics including asymmetric transmission, and asymmetric ellipticity and optical activity. Moreover, all these asymmetric effects are angle dependent, offering an additional degree of freedom for controlling the scattering and polarization properties of electromagnetic waves.

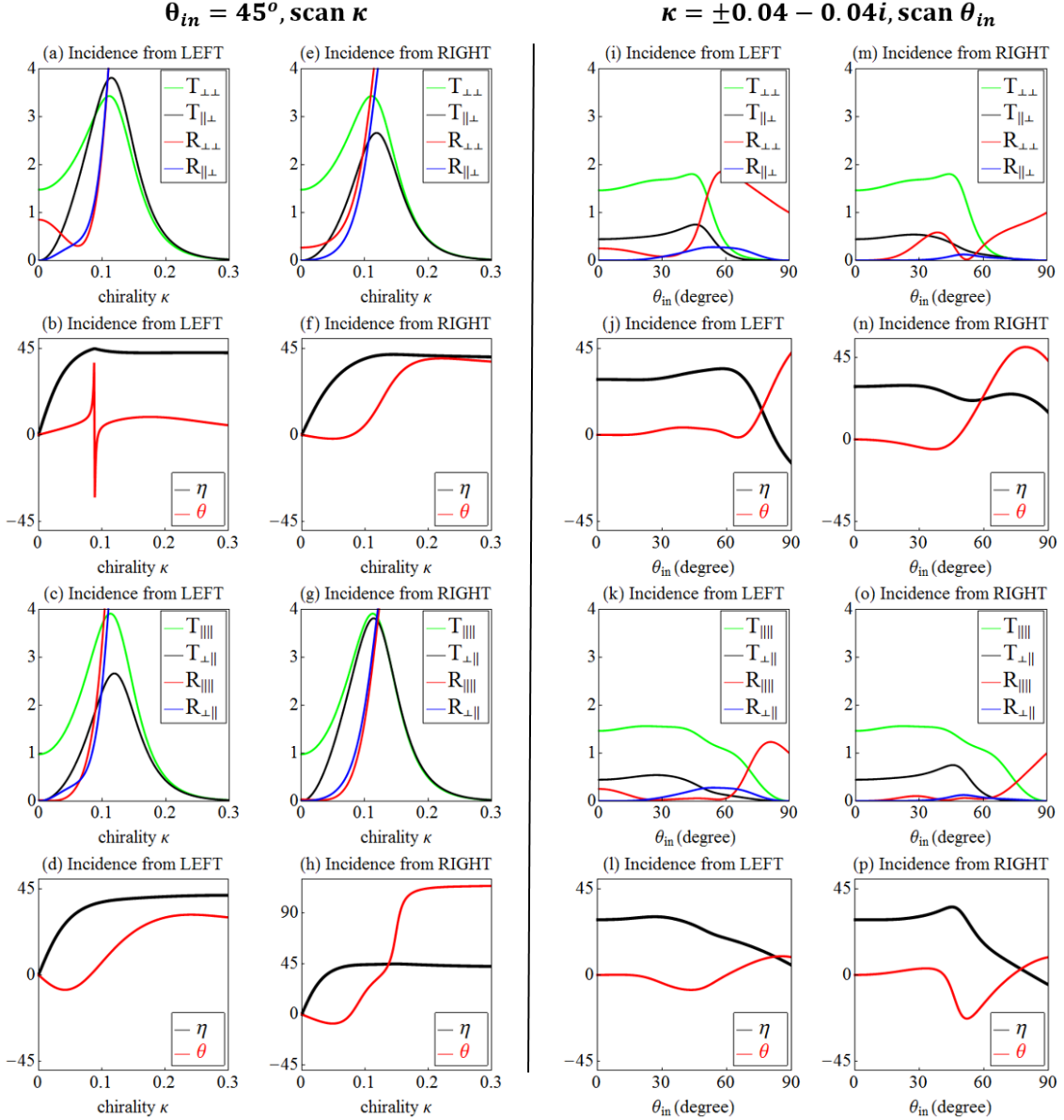

**Figure 6.** Reflection (*R*), transmission (*T*), optical activity (*θ*) and ellipticity (*η*) for a linearly polarized plane wave incident at both sides of a chiral bilayer as the one of Fig. 1, with permittivity and permeability values as in Fig. 2, at $\omega L/c = 15.5$ as a function of the chirality parameter (left two columns) and as a function of incident angle (right two columns). The subscripts ∥ and ⊥ in the transmission and reflection indicate the components parallel and perpendicular to the plane of incidence, respectively, while the first (second) component indicates the output (input) wave polarization. The coefficients *R* and *T* are the squared magnitude of the corresponding coefficients (*r* and *t*) of Eqs. (20), (21).

## 4. Conclusions

We investigated the effects of combining in the same system PT-symmetry (gain-loss media) and chirality. Starting by showing that such a combination is possible (despite the mirror-asymmetry of the chiral media) and by deriving the necessary conditions, we applied those conditions in a simple chiral bilayer and we investigated its different PT-related phases and its scattering- and polarization-related properties. In the case of normally incident electromagnetic waves we found that the PT-symmetric features, i.e. exceptional points, PT-phases etc., are not affected at all by chirality; as a result PT-symmetric and chiral effects can be controlled simultaneously and independently. In the case of oblique incidence, though, chirality highly affects the PT-symmetry related properties. In that case we showed that a full tuning of PT phases and scattering properties of our systems can be realized both by scanning the chirality parameter and/or the angle of incidence of the incoming waves. Moreover, we showed that by combining *PT*-symmetry with chirality we can achieve exotic propagation and scattering functionalities, such as asymmetric (side-dependent) transmission, ellipticity and optical activity. All the above effects can be exploited in a wide range of applications, in particular in applications where advanced polarization control is required (realizing, e.g., chiral unidirectional laser-absorbers, controllable polarization isolators etc.).

**Author Contributions:** M.K. and S.D. conceived the idea and designed the parametric study. C.M.S., E.N.E., and M.K. supervised the project. I.K. and S.D. carried out the analytical and numerical calculations. All authors discussed the results and commented on the manuscript, which was written by I. K..

**Funding:** This research was funded by the Hellenic Foundation for Research and Innovation (HFRI) and the General Secretariat for Research and Technology (GSRT), under the HFRI PhD Fellowship grant (GA. no. 4820), as well as by the EU-Horizon2020 FET projects Ultrachiral and Visorsurf.

**Acknowledgements:** The authors would like to thank Prof. K. Makris for useful discussions.

**Conflicts of Interest:** The authors declare no conflict of interest.